\documentclass[12pt,english]{article}
\usepackage{graphicx}
\usepackage{epsfig}
\usepackage[english]{babel}
\usepackage[ansinew]{inputenc}
\usepackage[T1]{fontenc}
\usepackage{latexsym}
\usepackage{amssymb}

\begin{document}
\title{Simulations of neutrino oscillations with a high-energy beta beam from CERN to LENA at Pyh\"asalmi Mine}
\author{Juha Peltoniemi\\ \em Excellence Cluster Universe \\ \em Technische Universität München, Garching, Germany}
\date{\today}
\maketitle

\begin{abstract}
I consider a high-Q beta beam peaking at multi-GeV energy for the baseline CERN-Pyh\"asalmi, with the length of 2288 km, using LENA, a 50 kton liquid scintillator as the far detector. The beta beam is assumed to be accompanied by a conventional wide band beam of 1--6 GeV. This combination turns out to be plausible to measure neutrino parameters if $\sin^2 2\theta_{13}\sim (1...3)\cdot 10^{-3}$.

\end{abstract}

\section{Introduction}

Beta beams have been suggested as a future option for long baseline neutrino experiments\cite{Zucchelli:2002sa,Mezzetto:2003ub,Lindroos:2003kp,:2008xx}. They provide a clean beam of electron (anti)neutrinos, and an alternative channel to conventional neutrino beams. Usually beta beams have been considered in connection with superbeams.

Here I study application of beta beams for the 2288 km long baseline from CERN to Pyh\"asalmi Mine in the middle of Finland. The density profile for the baseline has been modelled well \cite{Kozlovskaya:2003kk,Peltoniemi:2006hf}, and the average density can be defined at 1 \% accuracy. 

For the far detector I take LENA\cite{Wurm:2007cy, Oberauer:2006cd, Hochmuth:2006gz, MarrodanUndagoitia:2006qs, MarrodanUndagoitia:2006qn, MarrodanUndagoitia:2006rf, MarrodanUndagoitia:2006re, Undagoitia:2005uu, Autiero:2007zj}. The main purpose of LENA is to study low-energy neutrinos and nucleon decay, but it has been suggested that a large volume liquid scintillator indeed has capacity to measure high-energy neutrinos with a good energy resolution and flavor identification \cite{Peltoniemi:2009xx, Learned:2009rv}. Applying LENA for a low-energy (sub-GeV) beta beam was considered in \cite{marrodan}.   

Previously a conventional beam of 1--6 GeV from CERN to LENA at Pyh\"asalmi mine has been considered\cite{Peltoniemi:2009hv}. It turned out to be a very viable option to study neutrino parameters if the third mixing angle not too small, i.e. $\sin^2 2\theta_{13} >3\cdot10^{-3}$. Its range could be extended to smaller angles by increasing the detector size and beam power, but not very effectively because the performance is then limited by the beam background. 

The beta beam avoids the beam background. Hence it may provide a more effective way to extend the range to very small mixing angles. The viability of the beta beam, however, is still subject to substantive investigation.

\section{Beta beams}

\begin{figure}[tbp]
\begin{center}
\includegraphics[angle=0, width=10cm]{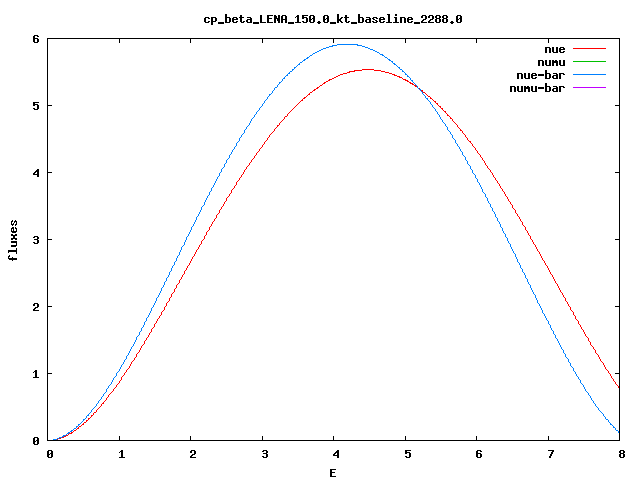}
\caption{\sf Spectrum of a beta beam, with $\gamma=300$. }
\label{betaspectrum}
\end{center}
\end{figure}

With the usual oscillation parameters the first oscillation peak of a 2288 km long baseline is at about 4.2 GeV. Hence an optimal beam covers the energies 1--5 GeV. 

To reach such energies I assume that beams of electron neutrinos and antineutrinos are produced from the decays of $\mbox{}^8$B and $\mbox{}^8$Li ions, with endpoints 14.43 MeV and 13.47 MeV, respectively \cite{Rubbia:2006pi,Rubbia:2006zv}. Using SPS we may reach $\gamma=350$.  As intensity I use the nominal value of $2\cdot 10^{18}$ decays per year, and $4+4$ years running time. Some people have been speculating on intensities up to $10\cdot 10^{18}$ decays per year, but it is not yet given that even the nominal intensity can be reached.

The spectrum of the beta beam is defined by $\gamma$, as shown in Fig.~\ref{betaspectrum}. In this work I set $\gamma=300$ which positions the peak at 4.2 GeV. Even lower $\gamma$, like 200 is still acceptable if it is more economic.

Use of $\mbox{}^{18}$Ne and $\mbox{}^6$He is disfavoured because of lower energies reachable (endpoint energies 3.5 MeV and 3.4 MeV, and peak energies 1.9 and 1.2 GeV with refurbished SPS). To reach the required 4--5 GeV energy we need $\gamma>1200$ which is possible only with LHC, though this possibility is supposedly ruled out due to some technical and availability limitations.   

I assume that the beta beam is accompanied by a wide band beam, as described in \cite{Peltoniemi:2009hv}.
The beta beam is mostly redundant to that, the main signal being $\nu_e \to \nu_\mu$ transition. It adds more statistics by increasing the beam power, but it has also an additional advantage to provide a clean beam with no beam contamination which is the bottleneck of the conventional beams for very low $\theta_{13}$.

\section{LENA}

LENA is planned to consist of 50 kton of liquid scintillator, in a vertical cylindrical tank with radius of 12 m and height of 100 m. It is surrounded by a buffer of 2 m, and outside it is shielded by water (totally 100 kton). The dimensions and number of phototubes may still vary a little until they are fixed soon.

Due to the detector geometry the efficiency to measure highest-energy muon neutrinos is reduced in the default vertical orientation. In these simulations I just assume the efficiency to decrease from 100 \% at 3 GeV to zero at 7 GeV. In a horizontal orientation the cut-off would be much milder.  

I consider also a larger size, 150 kton, for comparison. Additional detector volumes might be required to exploit the expensive beam optimally. On the other hand, for consistency I use throughout this work the detector mass as the scaling factor, containing also possible variations in the beam power, running time or the detector efficiency. One might note, however, that in real life the two beams may not scale equally.

Additional fiducial mass can be reached either with a single LENA by using the buffer and the shield as additional fiducial volumes --- the viability of which needs to be studied\footnote{The light-yield of the buffer is 2--3 magnitudes smaller than that of the scintillator, being mostly Cherenkov light. Nevertheless, because of small distance to the light sensor, $\sim 2$ m, it may be measurable, given also the immediate emission of the Cherenkov light, though only for outgoing particles. (The scintillation light, on the other hand, may be very delayed.) The outer shield, consisting of close to 100 kton of water, is going to be used as a Cherenkov detector to veto background muons, but would also serve as additional detector for tracks extending beyond the internal volume.} --- or with an additional 100 kton detector, like a horizontal high-energy scintillator aligned along the beam. Probably the combination with another experiment like a 100 kton liquid argon experiment GLACIER\cite{Rubbia:2009md} would give rather similar results. An additional detector dedicated for high-energy muons might indeed be very cost-effective.  

Increasing the beam power (which is a very unknown quantity) would have similar effect as increasing the detector size. Detector sizes even up to 300 kton (or equivalent beam empowerment) would not be completely unrealistic, if there is true need for that.

Throughout this work I assume the energy resolution of 3 \%, unless otherwise stated.
In real life the energy resolution may be a very complex function, and depends quite substantially on the properties of the scintillator, phototubes and electronics. The accuracy under different design factors of the detector is currently under a detailed study and more specific estimates will be released soon. The above assumption itself may not pose excessive requirements for the detection system, but the beam option has to be definitively taken into account on the design of the experiment to reach that. Studies with wide band beam with this beamline have shown that a 5 \% resolution is sufficient to study the CP violation and the mass hierarchy \cite{Peltoniemi:2009hv}.

I assume very good flavor identification. For the conventional beam, the main background is beam contamination. The beta beam is more background-free but I assume the background rate by neutral currents as 0.15 \% and the flavor misidentification rate  $10^{-4}$ for the electron neutrino disappearance channel.

\section{Results}

The simulations are done using the GLOBES codebase\cite{Huber:2007ji,Huber:2004ka}. The code is embedded within an own code.

\begin{figure}[tbhp]
\begin{center}
\includegraphics[angle=0, width=7cm]{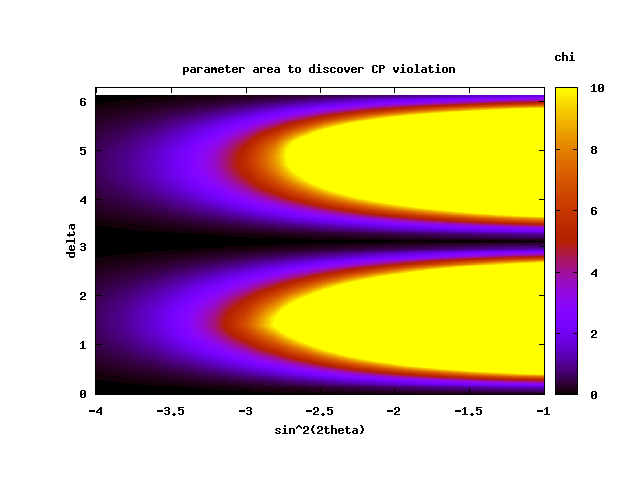}
\includegraphics[angle=0, width=7cm]{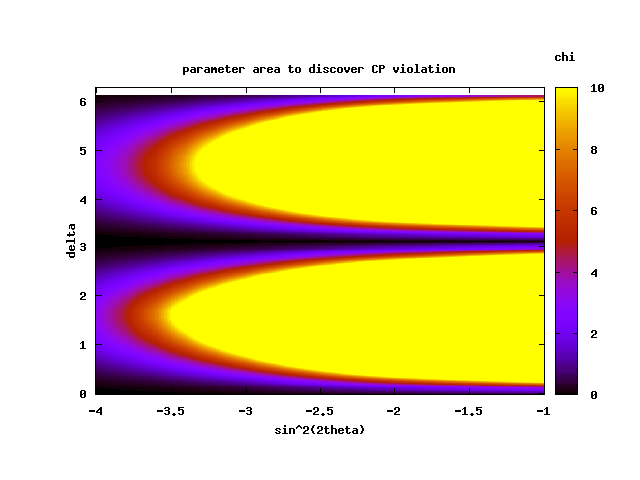}
\caption{\sf The expected range to measure the CP violation with a beta beam and a wide band beam together for a 2288 km long baseline. Left 50 kton and right 150 kton. There is a major difference, the larger is much better.
The colours represent $\chi$ values, so that yellow is about $3\sigma$, red $2\sigma$ and blue $1\sigma$.}
\label{cp_bw}
\end{center}
\end{figure}

\begin{figure}[tbhp]
\begin{center}
\includegraphics[angle=0, width=7cm]{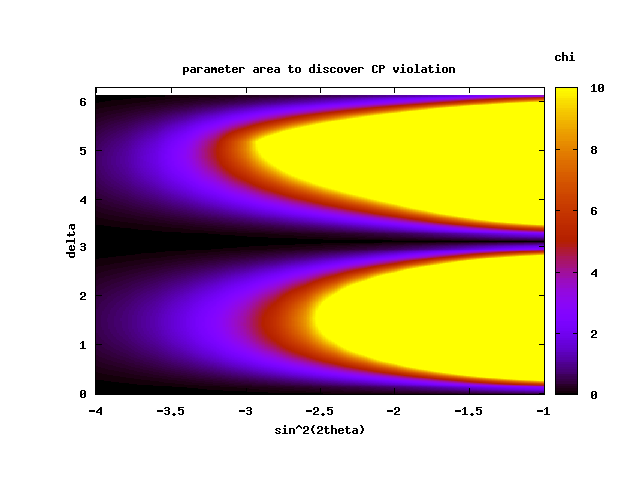}
\includegraphics[angle=0, width=7cm]{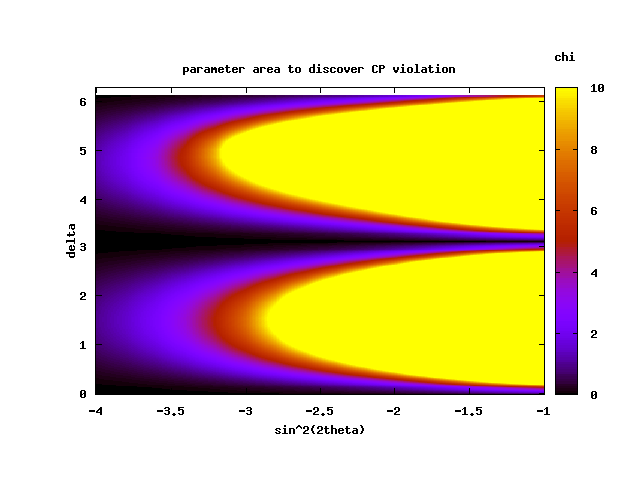}
\caption{\sf For comparison, the CP range with only a wide band beam. Left with 150 kton and right with 300 kton.}
\label{cp_w}
\end{center}
\end{figure}

\begin{figure}[tbhp]
\begin{center}
\includegraphics[angle=0, width=7cm]{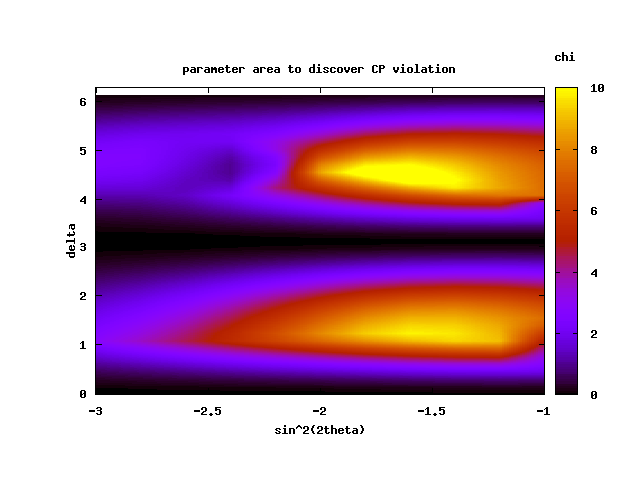}
\includegraphics[angle=0, width=7cm]{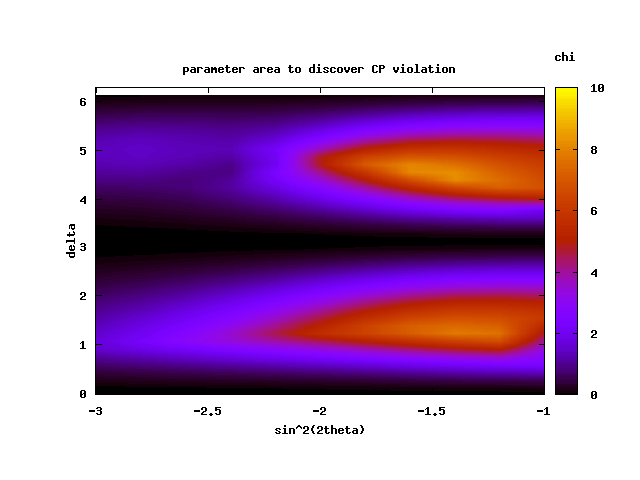}
\caption{\sf For comparison, the CP range with only a beta band beam with a 50 kton detector. Left for the horizontal layout (no high-energy cut) and righ for a vertical layout, i.e. reduced high-energy efficiency. The performance is very poor compared with the conventional beam. With beta beam, the orientation of the detector matters, a vertical alignment is a major burden.}
\label{cp_b}
\end{center}
\end{figure}

The observable range of the CP angle is obtained comparing $\delta=0$ or $\delta=\pi$ to different values and getting the respective $\chi^2$ values. The obtained $\delta$-plot is shown in Fig.~\ref{cp_bw}, with both 50 kton and 150 kton fiducial masses. At best it extends to $\sin^2 2\theta_{13} \sim 5\cdot 10^{-4}$ with 150 kton or $\sin^2 2\theta_{13} \sim 1.5\cdot 10^{-3}$ with 50 kton. This can be compared with the respective range of a wide band beam,  $\sin^2 2\theta_{13} \sim 1.5\cdot 10^{-3}$ with 150 kton or $\sin^2 2\theta_{13} \sim 8\cdot 10^{-4}$ with 300 kton, in Fig.~\ref{cp_w}. Also is shown that with only the beta beam (Fig.~\ref{cp_b}), to demonstrate that alone it is not good enough, with the assumed beam power, but it should be accompanied with a conventional beam (whose cost is anyway an order of magnitude less.)

\begin{figure}[tbhp]
\begin{center}
\includegraphics[angle=0, width=7cm]{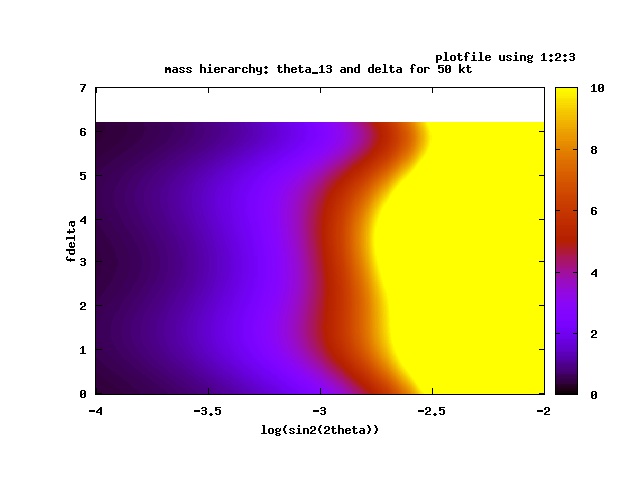}
\includegraphics[angle=0, width=7cm]{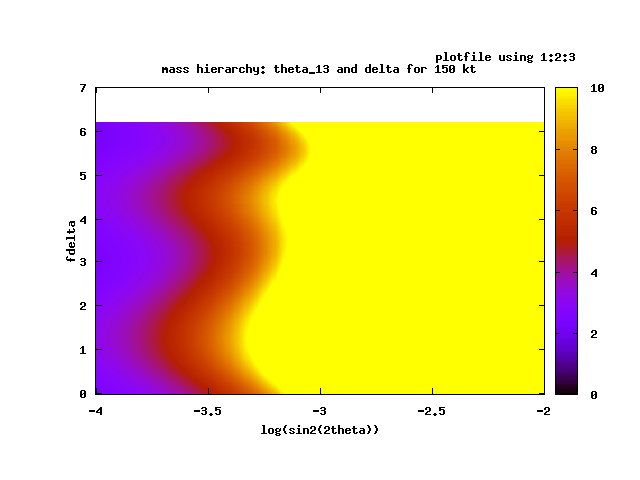}
\caption{\sf The capacity to measure the mass hierarchy in the plane of $\theta_{13}$ and $\delta$, with 50 kton and 150 kton detectors, both beams together.}
\label{mh_bw}
\end{center}
\end{figure}

\begin{figure}[tbhp]
\begin{center}
\includegraphics[angle=0, width=10cm]{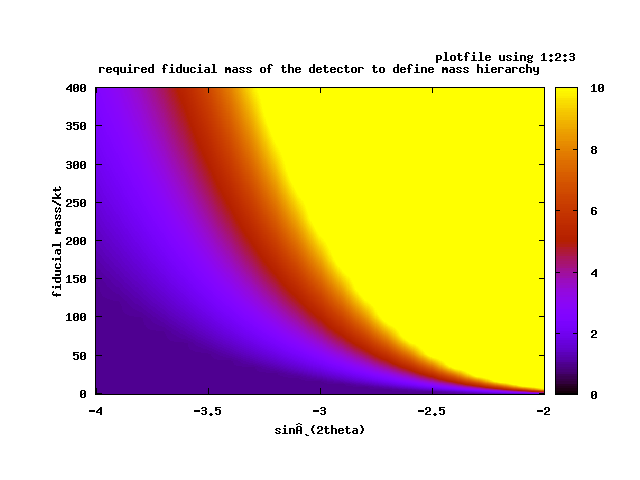}
\caption{\sf The capacity to measure the mass hierarchy with different fiducial masses of the detector, with a beta beam and a wide band beam together. The fiducial mass is used here as a general scaling variable. Here and in other similar plots $\delta=\pi/2$.}
\label{mhf_bw}
\end{center}
\end{figure}

The mass hierarchy is studied comparing events of $\delta m^2_{L}$ with $\delta m^2_{S}-\delta m^2_{L}$, marginalizing over other parameters. The beta beam improves the range at small mixing angles. In the given examples, with 150 kton detector $\sin^2 2\theta_{13} \sim 2\cdot 10^{-3}$ is reachable. With a wide band beam only, we would need a 400 kton detector to reach the same range. The beta beam alone does not, however, perform acceptably. Here we note that the behavior in CP angle is quite opposite for beta and wide band beams (assumed similar running times for both polarities), so they complete each others.

The range to measure $\theta_{13}$ was obtained comparing $\theta_{13}= 0$ to $\theta_{13} > 0$ when marginalising over other parameters. The range with beta beams extends to $\sin^2 2\theta \sim 3\cdot 10^{-4}$ or less. The combined beam with 100 kton detector will have a performance comparable to the wide band beam with 400 kton detectors.

\begin{figure}[tbhp]
\begin{center}
\includegraphics[angle=0, width=7cm]{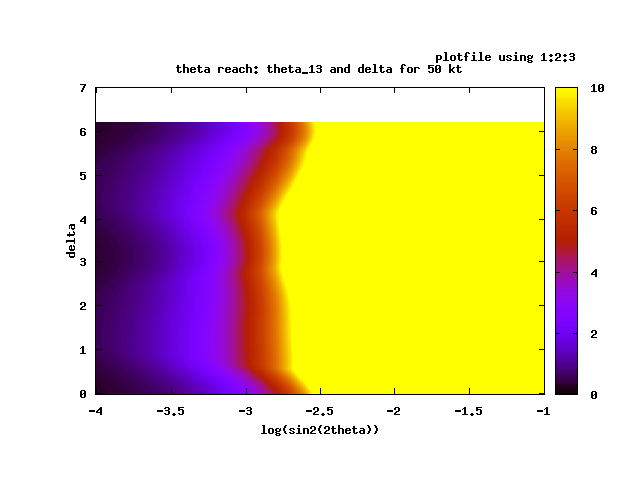}
\includegraphics[angle=0, width=7cm]{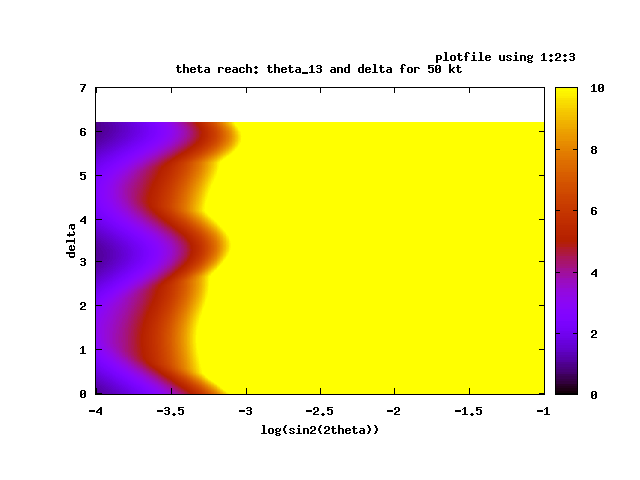}
\caption{\sf The reach of $\theta_{13}$ for 50 kton an 150 kton detectors with a beta beam and a wide band beam.}
\label{theta_bw}
\end{center}
\end{figure}

\begin{figure}[tbhp]
\begin{center}
\includegraphics[angle=0, width=10cm]{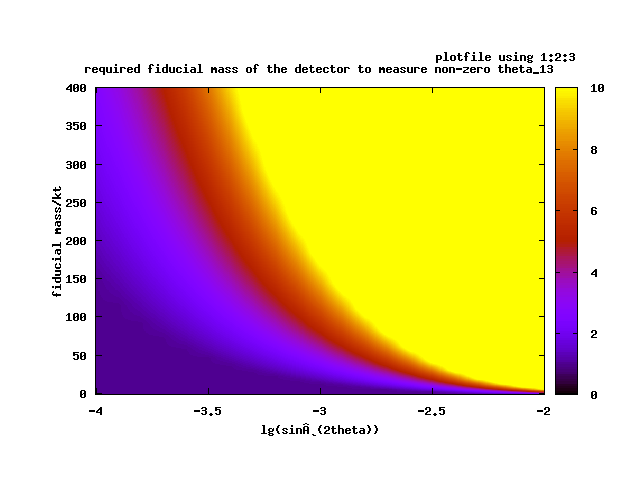}
\caption{\sf The reach of $\theta_{13}$ for fiducial masses of the detector with a beta beam and a wide band beam.}
\label{theta_f_bw}
\end{center}
\end{figure}

\begin{figure}[tbhp]
\begin{center}
\includegraphics[angle=0, width=7cm]{CPrange_bw_H150.png}
\includegraphics[angle=0, width=7cm]{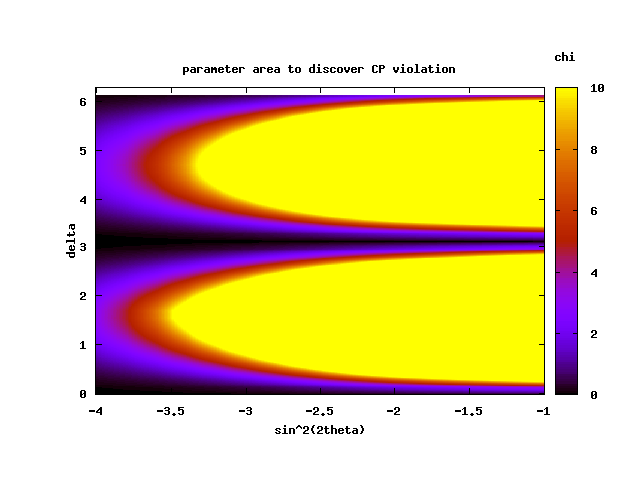}
\includegraphics[angle=0, width=7cm]{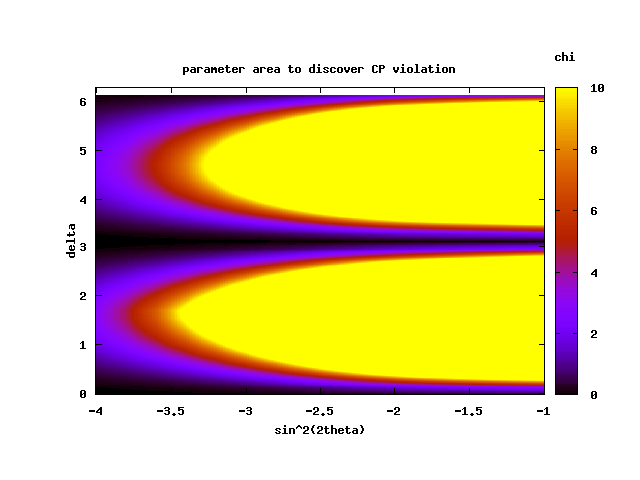}
\includegraphics[angle=0, width=7cm]{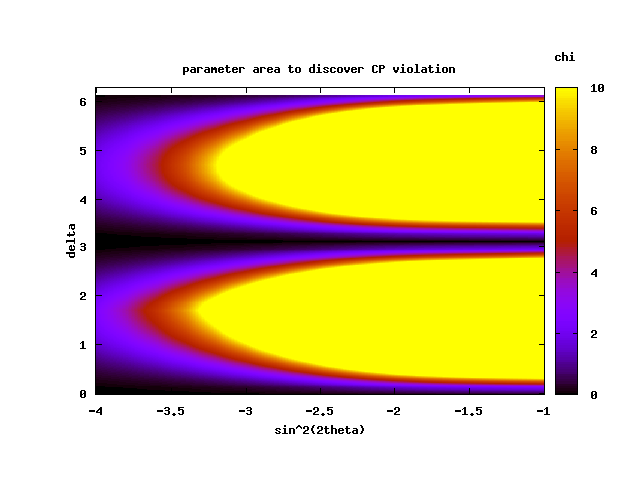}
\caption{\sf The CP range with both beams combined and a 150 kton detector, with 3 \%, 5 \% 10 \% and 20 \% energy resolutions. The last one is visibly worse, but the difference between the first ones is rather marginal. Hence 5 \% resolution would be sufficiently optimal while resolutions up to 10 \% would still be accaptable.}
\label{cp__bw_rel}
\end{center}

\end{figure}

\begin{figure}[tbhp]
\begin{center}
\includegraphics[angle=0, width=7cm]{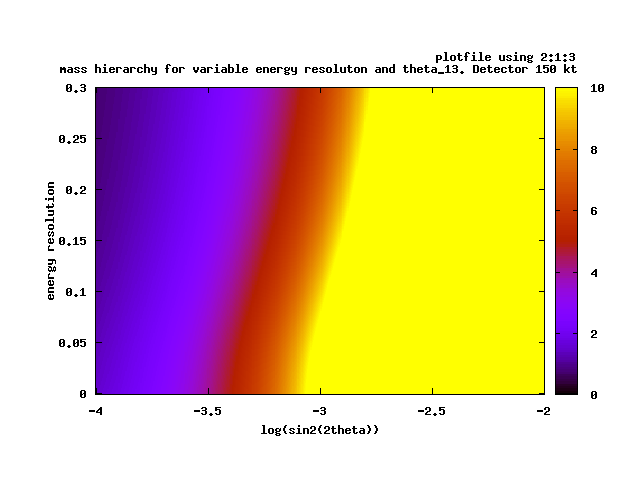}
\includegraphics[angle=0, width=7cm]{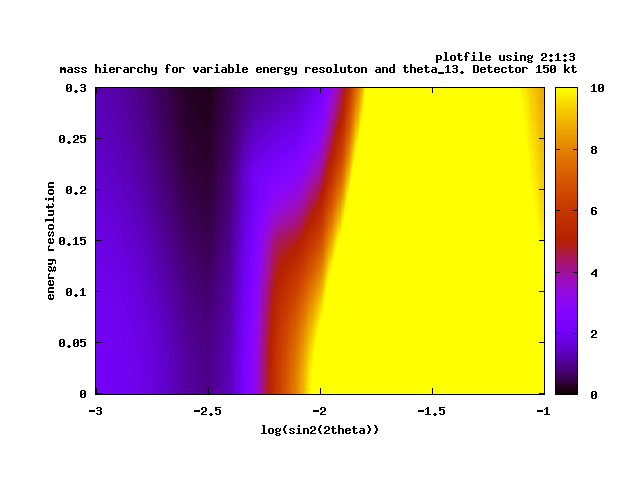}
\caption{\sf The capacity to measure the mass hierarchy with different relative energy resolutions, for both beams together (left) and for only beta beam (right).}
\label{eres}
\end{center}
\end{figure}

The requirements for the energy resolution are rather mild. As seen in Figures \ref{cp__bw_rel} and \ref {eres}, the resolution 5 \% is certainly good enough but even worse resolutions up to 10 \% are still acceptable. The requirements are not stronger than in the case of conventional beam, rather weaker.

\section{Conclusions}

The beta beam increases the capacity to measure the neutrino properties with small mixing angles. The beta beam allows to study $\sin^2 2\theta_{13} \sim 10^{-3}$ with the postulated setup with 50 kton. This is subtantially better than that reachable with a conventional wide band beam with similar detectors at the same baseline, depending though on the available beam powers. 

For the energy resolution 5 \% was found to be sufficient. This should be taken as the design goal for LENA, though even better resolutions may be achievable. A monoenergetic beta beam would not bring any advantage with LENA.

Increasing the power of the beta beam, or enlargening the detector improves the capacity accordingly, because the performance is not limited by beam systematics. With a 150 kton detector, or a respectively stonger beam, mixings up to $\sin^2 2\theta_{13} \sim (3...5)\cdot 10^{-4}$ are studiable. Given the high cost of the beam, larger detectors would be desired to balance the overall investment. Probably the optimal fiducial detector size for the beta beam of considered power would be at least 100 kton. Also, the standard LENA with another detector like GLACIER might perform rather equally.

For the beta beam the vertical orientation of the tank imposes a small penalty on the performance as many of the  longest muon tracks are not contained within the fiducial volume. A horizontal alignment would be hence preferable, but the problem could be also partially circumvented by using the outer buffer and shield to measure the tails of the outgoing muon tracks.
\\

{\bf Acknowledgement}\\
This research was supported by the DFG cluster of excellence "Origin and Structure of the Universe". I thank the LENA collaboration for inspiring co-operation, advice and help, E15 group of Franz von Feilitzsch for hospitality and the LAGUNA consortium.



\begin{figure}[tbhp]
\begin{center}
\includegraphics[angle=0, width=7cm]{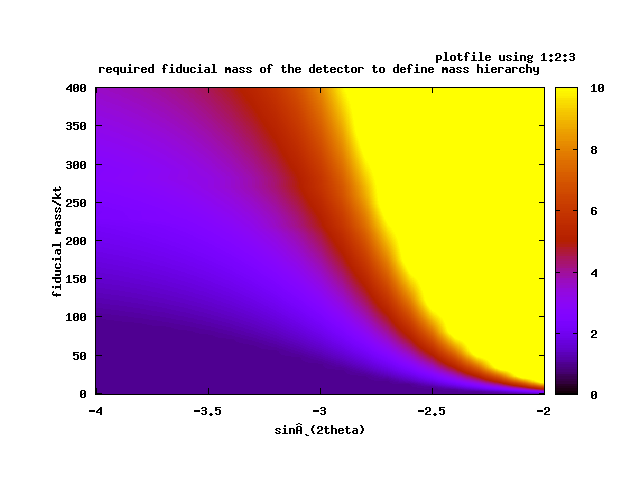}
\includegraphics[angle=0, width=7cm]{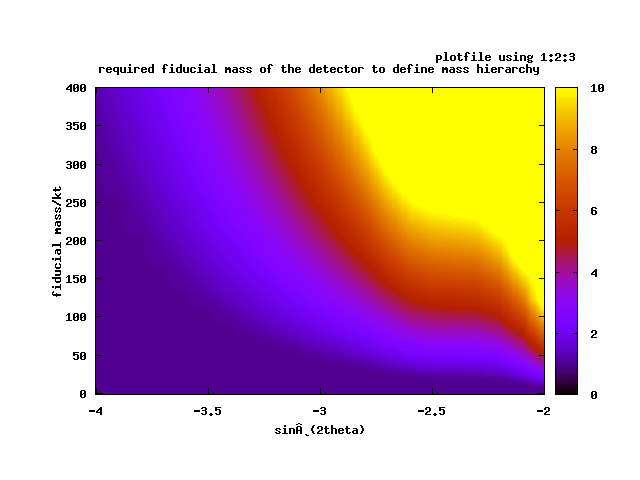}
\caption{\sf For comparison, the capacity to measure the mass hierarchy with the detector masses, with only one beam, a wide band beam (left) and a beta beam (right).}
\label{mhf_w}
\end{center}
\end{figure}

\begin{figure}[tbhp]
\begin{center}
\includegraphics[angle=0, width=7cm]{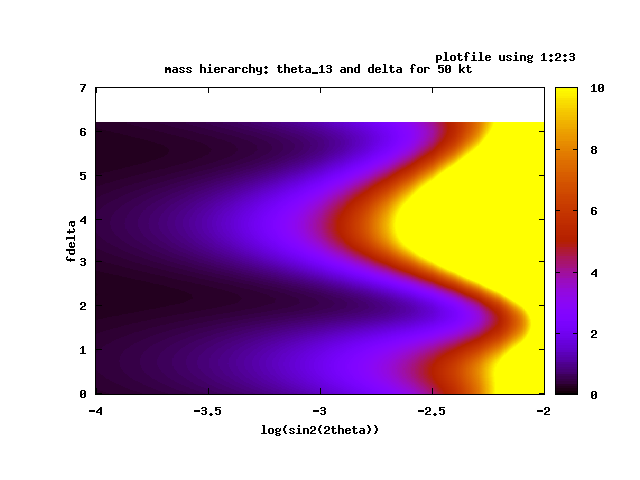}
\includegraphics[angle=0, width=7cm]{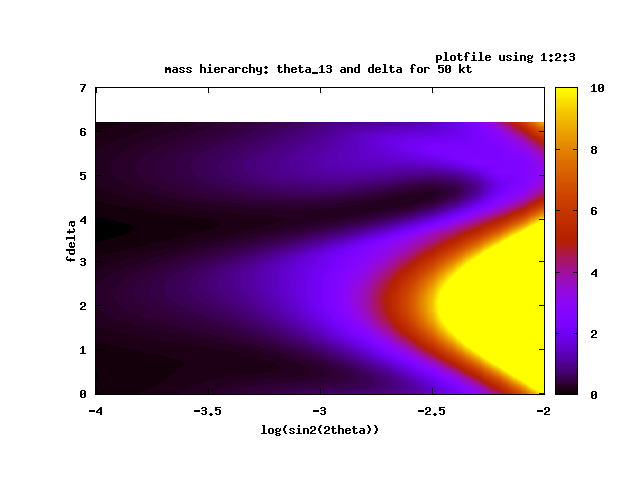}
\caption{\sf For comparison, the hierachy plot for a 50 kton detector with only a wide band beam (left) and only beta beam (right).}
\label{mh_50}
\end{center}
\end{figure}

\begin{figure}[tbhp]
\begin{center}
\includegraphics[angle=0, width=7cm]{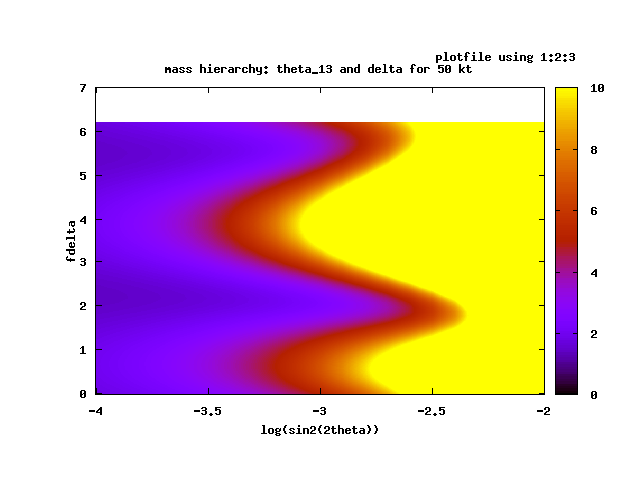}
\includegraphics[angle=0, width=7cm]{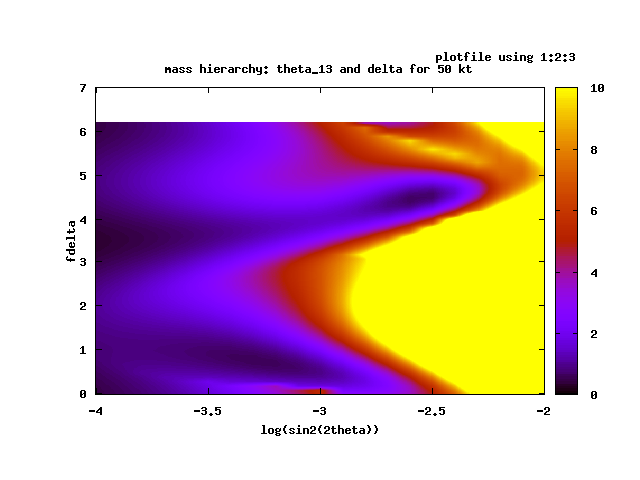}
\caption{\sf For comparison, the hierarchy plot for a 150 kton detector with only a wide band beam (left) and only beta beam (right), 150 kton.}
\label{mh_150}
\end{center}
\end{figure}

\begin{figure}[tbhp]
\begin{center}
\includegraphics[angle=0, width=7cm]{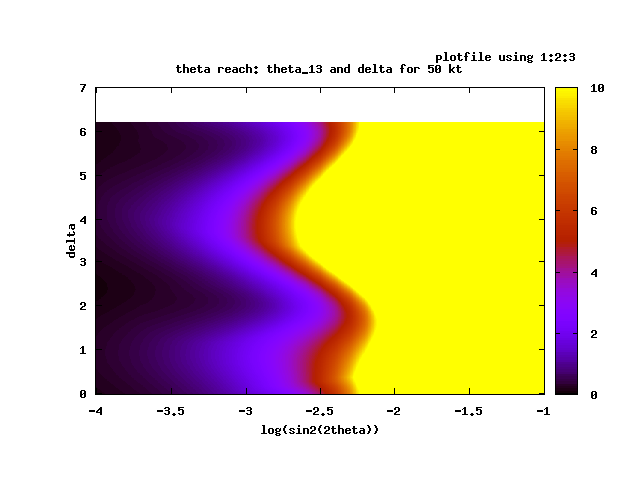}
\includegraphics[angle=0, width=7cm]{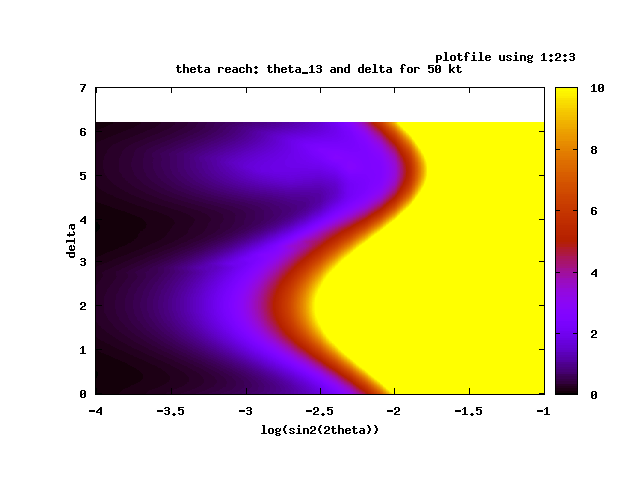}
\caption{\sf For comparison, the reach in $\theta_{13}$ with only a wide band beam (left) and only a beta beam (right), 50 kton detector. Note also the different behavior with respect to the CP angle.}
\label{theta_w}
\end{center}
\end{figure}

\begin{figure}[tbhp]
\begin{center}
\includegraphics[angle=0, width=7cm]{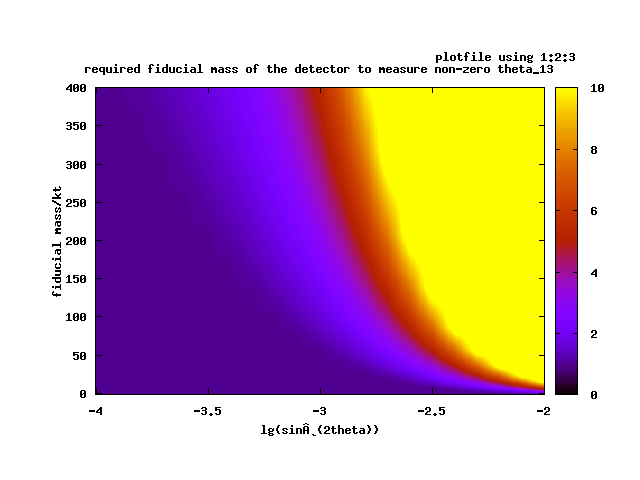}
\includegraphics[angle=0, width=7cm]{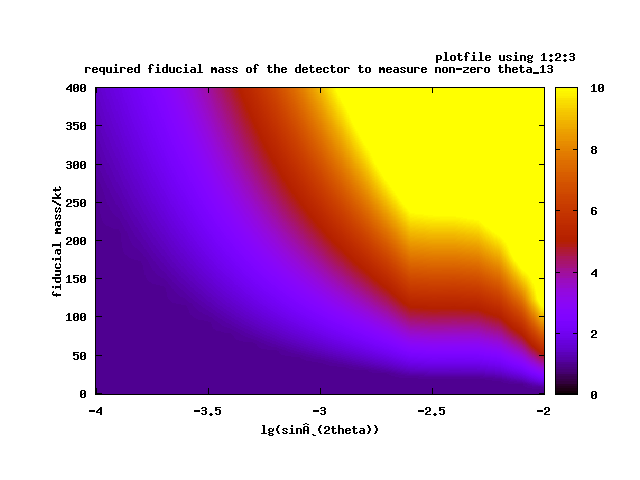}
\caption{\sf For comparison, the range in $\theta_{13}$ measurements related to the fiducial mass of the detector with only a wide band beam (left) and only a beta beam (right).}
\label{theta_f_w}
\end{center}
\end{figure}


\end{document}